\documentclass[useAMS,usenatbib]{mn2e}
\usepackage{graphicx}
\usepackage{xcolor}

\newcommand\ion[2]   {#1\,{\sc #2}}
\newcommand {\nii}   {[\ion{N}{ii}]}
\newcommand {\oiii}  {[\ion{O}{iii}]}
\newcommand {\oi}    {[\ion{O}{i}]}
\newcommand {\sii}   {[\ion{S}{ii}]}
\newcommand {\heii}  {\ion{He}{ii}}

\newcommand {\ha}    {\ifmmode ${H}$\alpha \else {H}$\alpha$\fi}
\newcommand {\hb}    {\ifmmode ${H}$\beta  \else {H}$\beta$\fi}
\newcommand {\hg}    {\ifmmode ${H}$\gamma \else {H}$\gamma$\fi}
\newcommand {\laa}   {\ifmmode ${Ly}$\alpha \else {Ly}$\alpha$\fi}
\newcommand {\ergs}  {\ifmmode ${\ergs}$ \else ergs\,cm$^{-2}$\,s$^{-1}$\fi}
\newcommand {\ergsA} {\ergs\,\ifmmode ${\AA}$^{-1} \else \AA$^{-1}$\fi}
\newcommand {\kms}   {\ifmmode $km\,s$^{-1} \else km\,s$^{-1}$\fi}

\newcommand\tabref[1] {Table~\ref{#1}}
\newcommand\figref[1] {Fig.~\ref{#1}}

\title[Spectral variability of 3C~390.3]{Spectral variability of the 3C~390.3 nucleus for more than 20 years -- II. Variability of the broad emission-line profiles and \heii\,$\lambda$4686\AA\ emission-line fluxes}
\author[S. G. Sergeev]{S.~G.~Sergeev\thanks{E-mail: sergeev.crao@mail.ru}
\\
Crimean Astrophysical Observatory, P/O Nauchny Crimea 298409, Russia}

\begin{document}


\pagerange{\pageref{firstpage}--\pageref{lastpage}} \pubyear{}

\maketitle

\label{firstpage}

\begin{abstract}

Results of the analysis of the variability of the \hb\ and \ha\
broad emission-line profiles and the \heii\,$\lambda4686$\,\AA\ emission-line
fluxes in the 3C~390.3 nucleus during 1992--2014 are present.
The observed velocity-dependent lag for the Balmer lines
is similar to that expected from the Keplerian disc configuration,
although there are some differences.
Probably, a radial infall motion can be present in the broad-line region
of 3C~390.3 in addition to the Keplerian rotation.
The lag of the broad \heii\ line is $26\pm 8$\,d,
significantly less than that of the Balmer lines,
so the \heii\ emission region is much smaller in size.
In terms of the power-law
relationship between line and optical continuum fluxes with slowly
varying scalefactor $c(t)$: $F_{line}\propto c(t)\,F_{cont}^a$, the
power $a$ is $1.03$ for the broad \heii\ line, while according to \citet{Ser17}
the power is equal to $0.77$ and $0.54$ for the broad \hb\ and \ha\ lines, respectively.
It means that the variability amplitude is
the largest in the \heii, less in \hb, and more less in \ha.
However, the Balmer lines contain a long-term trend that is not
seen in the helium line.
The narrow \heii\ line is variable with
the amplitude (max-to-min ratio) $R_{max}\approx 3$ that is
much greater than
the variability amplitudes of both the narrow Balmer lines
and the narrow \oiii\,$\lambda$5007\,\AA\ line.

\end{abstract}


\section[]{Introduction}

A major goal of extragalactic researches is to understand spatially unresolved
internal structure of the active galactic nuclei (AGN), in particular, the
structure, kinematic, and physical conditions in the broad-line region
(BLR). The BLR consists of high velocity gas (up to $\sim$20000 \kms)
that is ionized and heated by the time-variable, high-energy continuum
of the central source. This gas produces broad emission lines (BEL).
While most of the BEL profiles are roughly symmetric, some AGNs
show strong redward or blueward asymmetries, bumps, and shelves
in the BEL profiles.
In addition, the BEL profiles are strongly variable
in flux, responding or ``reverberate'' to the continuum variations,
as well as in shape.

A long-term monitoring of AGNs is an effective way to study the BLR and
to distinguish among various proposed BLR models. High temporal resolution
time series of the continuum and broad-line fluxes are of particular interest
for the reverberation mapping studies (RM). RM \citep{Bla82,Pet93,Pet14}
is a powerful tool to investigate the BLR size, structure,
and kinematics. Under some simple assumptions, the emission-line
response to the continuum variations is a convolution of the
continuum light curve with a so-called transfer function $\Psi(\tau)$
(for the total line flux) or $\Psi(\tau,\lambda)$ (for the line flux at
wavelength $\lambda$).
The transfer function reflects BLR size, geometry, and kinematics.
In its simplest form, the mean time delay  (or lag $\tau$)
between continuum and emission-line variations is measured
by the cross-correlation function (CCF) for the respective light curves and it
represents a typical BLR size. The lag at the CCF peak
(or $\tau_{peak}$) is approximately equal to a lag that divides a transfer
function into two equal areas \citep{Ser99b}.

From the time delay as a function of line-of-sight velocity
it is possible to distinguish among geometric and kinematic BLR configurations
(e.g. flattened versus spherical geometry and rotation, infall, or outflow kinematics).
For example, the Keplerian disk produces a
velocity-symmetric structure with lower lags at the far profile wings
while the radial motion produces an asymmetric velocity structure
with smaller delays on the red/blue profile wings for infall/outflow, respectively.
Such detailed velocity-resolved analyses have been applied
to about ten AGNs \citep[e.g,][]{Ser99a,Den09,Ben10,Den10,Dor12,Gri13,Pei17,Rosa18}.
In general, the BLR kinematics has been found to be a rotational or infall motion,
but the BLR appears to be more complex in some objects,
e.g. combined virial motion and infalling gas in Mrk~6
\citep{Ser99a,Dor12,Gri13}.

However, the long-term profile variations (months to years)
seem to be completely independent of the continuum variations,
so their origin can not be ascribed to reverberation effects
\citep[e.g.][]{Ser02}.
Reasons for the long-term profile evolution are not clear.
They can be related to the multi-component BLR model with
multiple physical components \citep[e.g.][]{Sti88,Pet90,Mal97,Ser01},
or to the inhomogeneities in a disk-like BLR such as
hot spots \citep{Zhe91,New97,Ser00,Flo08} or spiral arms \citep{Cha94,Sto17}, or to the
line emissivity redistribution in space that depends on the continuum brightness
\citep[e.g. broad \hb\ line was found to be broader when the continuum flux is weaker in
accordance with the photoionization models, e.g.][]{Cac06},
or to dynamic changes of the BLR, or to the anisotropic continuum emission
\citep[e.g.][]{Goad96}, or to some other reasons.

The nucleus of the radio galaxy 3C~390.3 is a prototype
to so-called double-peaked emitters, objects with strongly
double-peaked broad-line profiles (red and blue bumps or shoulders).
This nucleus is among the best-studied objects of this class
and its variability history is well-known
\citep[e.g.][]{Ost76,Bar80,Yee81,Net82,Vei91,Zhe96,Wam97,Die98,OBr98,Gez07,Sha10,Die12,Afa15}.

The double-peaked profiles are often believed to be
a signature of the relativistic Keplerian disk, either circular or elliptical
\citep[e.g.][]{ch89a,ch89b,Era95}.
Inhomogeneities in the disk surface brightness have been proposed
in order to explain variations of the profile shapes (see above).
Alternative BLR models to account for the double-peaked profiles include
outflowing biconical gas streams \citep{Vei91,Zhe91}
and binary black holes \citep[e.g.][]{Pet87,Gas96,Zha07}.

The present study is a continuation of the previous studies of the 3C~390.3
nucleus carried out in the Crimean Astrophysical Observatory (CrAO)
since 1992: \citet{Ser02,Ser11} and Sergeev, Nazarov \& Borman (2017)
\citep[hereafter][]{Ser17}.
The observations
are described in Sect.~\ref{sect-obs}. The velocity-resolved variability
characteristics and the velocity-resolved lags for the \hb\ and \ha\ lines
are presented in Sect.~\ref{sect-var} and \ref{sect-lag}. In Sect~\ref{sect-heii},
the lag for the \heii\,$\lambda$4686\,\AA\ broad emission-line is determined
and its variability patterns are compared to those of the Balmer
lines.
The obtained results are summarized in Sect~\ref{sect-sum}.

\section[]{Observations and data reduction}
\label{sect-obs}
\subsection[]{Optical spectroscopy}

Optical spectra of 3C~390.3 have been obtained at the 2.6-m Shajn telescope
of the CrAO since 1992. The results of the observations for the periods
1992--2000, 2001--2007, and 1992--2014 are published in \citet{Ser02,Ser11,Ser17},
respectively. The spectra were registered at the two
separate spectral regions centered at the H$\alpha$ and H$\beta$ lines.
More details about observations, observational setup,
and spectral data processing are in \citet{Ser02,Ser11,Ser17}. Below I briefly
recall how the spectra of 3C~390.3 were calibrated in flux.

At first, the spectra have been scaled to match the
fluxes of the selected narrow emission lines which are assumed to be constant
over time-scales of the monitoring programme.
The scaling procedure has been performed according to the method of residuals
as described by \citet{Gro92}.
However,
in the case of 3C~390.3 the underlying difficulty is that there are evidences for
the narrow-line variability \citep{cl87,Zhe95,Ser17}.
A discrepancy between spectral and photometric light curves of 3C~390.3
\citep{Ser17} has been attributed to this variability.
So, it was assumed that the photometric
measurements are correct, while the spectral measurements are wrong
because this variability. Therefore, in \citet{Ser17}
the spectra were re-scaled to achieve an agreement between the spectral and
photometric data sets. To correct the spectral measurements,
the $\varphi(t)$ function has been recovered to re-scale the spectra
as described in \citet{Ser11}.
It was selected to be a small-degree polynomial function
of time \citep[see][]{Ser17}.

The narrow lines have been separated
from the broad-line profiles
as described in
\citet{Ser02,Ser07}.
So, low-state spectra of 3C~390.3 were used to isolate
the narrow-line components. Unfortunately, in contrast to \citet{Ser07},
the \oiii\,$\lambda$5007 line is apparently blended with the broad \hb\ line.
Therefore, as a first iteration,
it was adopted that there is no broad \hb\ line underneath
the \oiii\,$\lambda$5007 line profile, but only a continuum
with a constant flux. This continuum
was measured at the red side of the \oiii\ profile and then subtracted from
\oiii.
Taking in a mind that the lines of the \oiii\,$\lambda4959+5007$ doublet
are scaled and shifted versions of each other,
an initial \oiii\,$\lambda$4959 profile can easily be obtained.
After subtraction of this profile,
the red \hb\ wing on the blue side of the \oiii\,$\lambda$5007 profile is free of
narrow lines,
and we can use both the blue and red wavelength windows for a local
straight-line pseudo-continuum that represents the red wing of the broad
\hb\ line beneath the \oiii\,$\lambda$5007 line.
After the third iteration, there are no more changes in the
obtained \oiii\,$\lambda$5007 narrow-line profile.

The obtained \oiii\,$\lambda$5007 profile has been used as a template
to separate other narrow lines (\hb, \heii, \ha, \nii, \sii, \oi), again for the
low-state \hb\ and \ha-region spectra.
The separation of the narrow lines was based on the
assumption that a broad-line profile beneath the narrow
line(s) is sufficiently smooth for the low-order interpolation underneath
the narrow line(s) and that the narrow-line profiles are similar to
that of \oiii\,$\lambda$5007.
For this purpose, the multi-dimensional optimization
algorithm was used.
Finally, the spectrum of the narrow lines has been created in order
to obtain broad-line components by subtraction the narrow-line components
from each spectrum.

The subtraction of the narrow lines
has been applied to the spectra that were scaled in flux using the
narrow-line fluxes 
\citep[\oiii\ and \oi\ in the \hb\ and \ha\ spectral
regions, respectively, see][]{Ser17}.
Since the narrow Balmer lines vary in flux stronger than that of
\oiii\ and \oi\ \citep{Ser17},
these lines were subtracted alone.
Then the narrow-line free \hb\ and \ha\
broad profiles were re-scaled using the $\varphi(t)$ scalefactor
to account for the narrow-line variability.

\begin{figure}
  \includegraphics[width=\columnwidth]{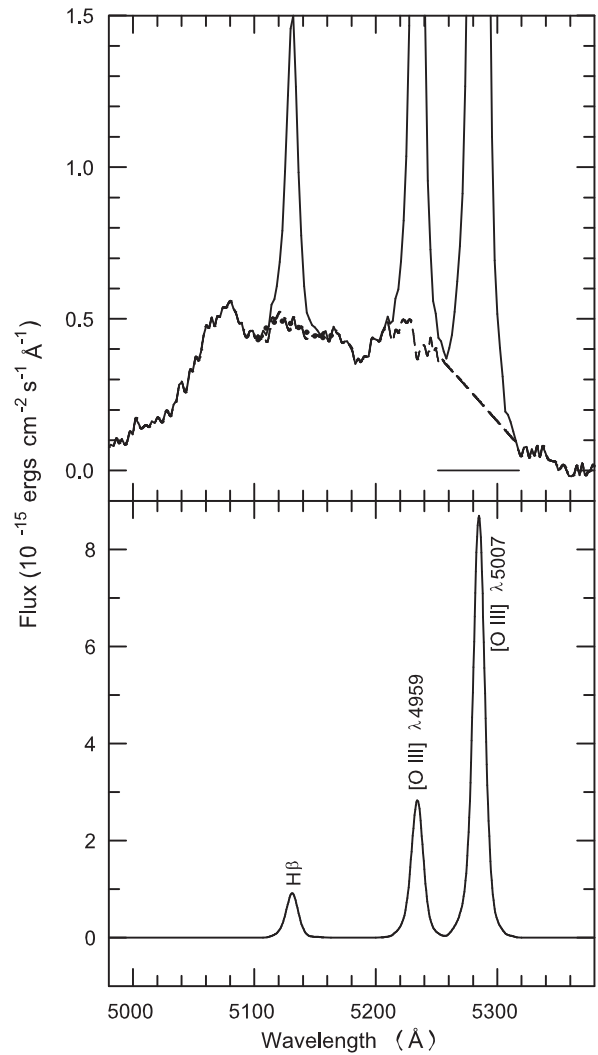}
\caption{
Remake and visualization of the separation of the narrow-line
components in the \hb-region in 3C~390.3, which was performed by \citet{Ser02}.
Top panel: Solid line is a mean of the low-state spectra, dashed line is
the \hb\ broad-line profile, dotted line is
the low-order interpolation underneath the \hb\ narrow-line profile
(see text for more details).
Horizontal line indicates the wavelength window for which
the red wing of the broad \hb\ line was assumed to be a straight line.
Bottom panel:
Narrow-line component that was used to obtain the \hb\ broad-line profile
for each spectrum.}
  \label{narrow}
\end{figure}

The \figref{narrow} illustrates the separation of the narrow lines
from the low-state mean spectrum, which was done by \citet{Ser02}.

While uncertainties in our H$\alpha$-region fluxes are much greater than in the
H$\beta$-region ones because large uncertainties
in the flux-scaling factors determined from the relatively
weak \oi\,$\lambda$6300 narrow line,
the flux-independent profile shapes of H$\alpha$ are more reliable
than that of H$\beta$.

\subsection[]{Optical photometry}
\label{phot}

Regular CCD broad-band photometric observations of the selected AGNs, including
3C.390.3, have been started at the CrAO in 2001.
The instrumentation, reductions and measurements of our photometric data are
described in \citet{dor05} and \citet{se05}.
The {\it V\/}-filter photometric measurements of 3C~390.3
were calibrated to match the $F_{5100}$ spectral fluxes of the continuum
and to obtain a joined continuum light curve \citep[see][for more details]{Ser17}.

\section[]{Results}

\subsection[]{Light curves for individual profile segments}

\begin{figure*}
  \includegraphics[width=120mm]{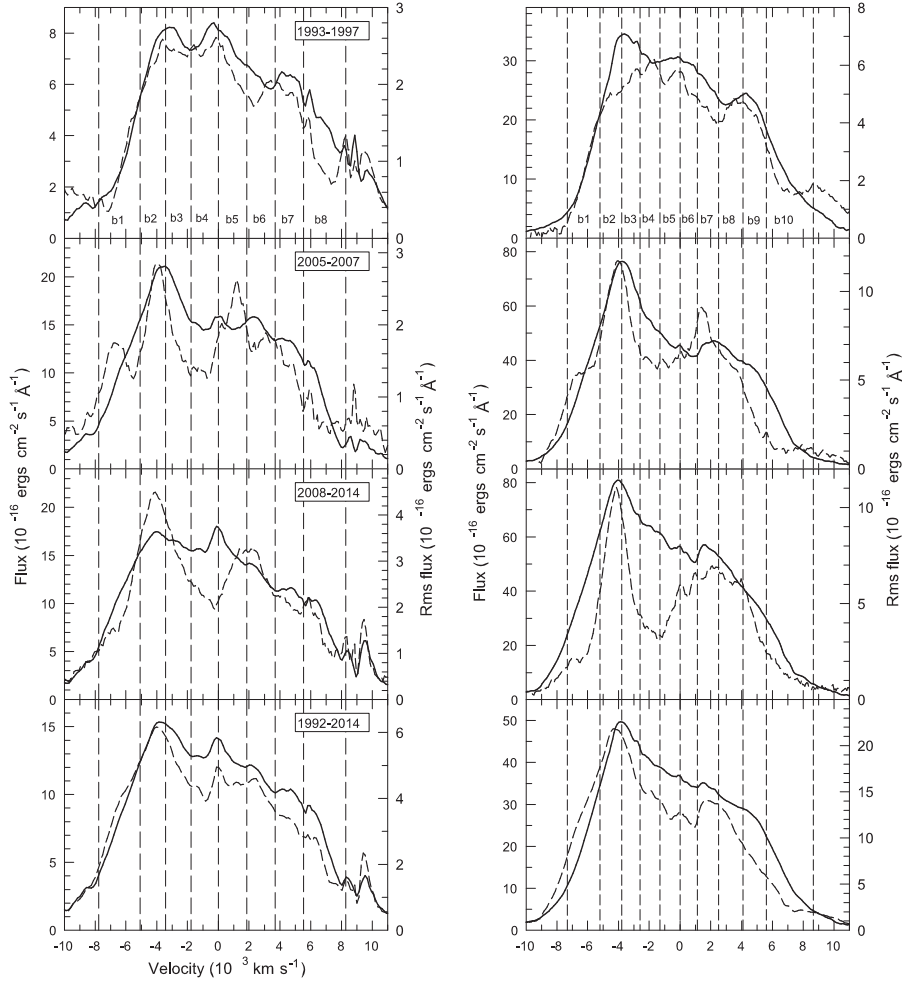}
\caption{Mean (solid lines) and rms (dashed lines) profiles of the broad \hb\
(left-hand panels) and \ha\ (right-hand panels) lines for the following
periods of observations (top-to-bottom): 1993--1997, 2005--2007, 2008--2014,
and 1992--2014.
The left-hand axes in each plot are for the mean fluxes, while the right-hand
axes in each plot are for the rms fluxes.}
  \label{mean-rms}
\end{figure*}

The \hb\ and \ha\ broad-line profiles were divided into eight
and ten velocity-space bins, respectively.
The bin boundaries are given in \tabref{tab-bins} and shown
in \figref{mean-rms}. The boundaries were chosen to have
approximately the same flux for each bin for the mean
as well as for rms profiles and by considering the four periods of observations
from \citet{Ser17}:
1993--1997, 2005--2007, 2008--2014,
and the entire period of CCD observations of 3C~390.3 at the CrAO (1992--2014).
The motivation for the data division into the periods is that
these periods were used for the cross-correlation analysis in
\citet{Ser02,Ser11,Ser17}. The 1998--2004 period has been excluded
from the cross-correlation analysis because
poor data sampling for both the \hb\ and \ha\ lines.
On the other hand, this division can be used to check for changes
in the variability characteristics, e.g. lag changes.
The light curves for the chosen bins were computed as given in
\citet{Ser02,Ser11}.

\begin{table}
\caption{Emission-line bins}
\label{tab-bins}
\begin{tabular}{lccc}
\hline
Bin  & Line & Boundaries  & Velocity      \\
No.  &      & (\AA)       & (\kms)\\
\hline
\hline
1  & \hb  & 5000--5045     & $-6424$ \\
2  & \hb  & 5045--5073     & $-4254$ \\
3  & \hb  & 5073--5101     & $-2599$ \\
4  & \hb  & 5101--5131     & $-886 $ \\
5  & \hb  & 5131--5163     & $+925 $ \\
6  & \hb  & 5163--5195     & $+2774$ \\
7  & \hb  & 5195--5227     & $+4620$ \\
8  & \hb  & 5227--5275     & $+6912$ \\
1  & \ha  & 6760--6808     & $-6253$ \\
2  & \ha  & 6808--6840     & $-4491$ \\
3  & \ha  & 6840--6867     & $-3198$ \\
4  & \ha  & 6867--6897     & $-1954$ \\
5  & \ha  & 6897--6927     & $-650 $ \\
6  & \ha  & 6927--6953     & $+562 $ \\
7  & \ha  & 6953--6985     & $+1812$ \\
8  & \ha  & 6985--7022     & $+3292$ \\
9  & \ha  & 7022--7058     & $+4850$ \\
10 & \ha  & 7058--7130     & $+7140$ \\
\end{tabular}
\end{table}

The narrow-line variability leads
to the appearance of fictive long-term trends in the
derived light curves when the obtained spectra are scaled in flux
using narrow-line fluxes.
All the results of the present paper are given
with accounting for this trend \citep[see][for more details]{Ser17}.

\subsection[]{Variability characteristics}
\label{sect-var}

The basic variability characteristics of 3C~390.3 for the
selected bins of the \hb\ and \ha\ broad-line profiles are summarized in
\tabref{var}. As in \citet{Ser17}, it was considered the four periods of observations:
1993--1997, 2005--2007, 2008--2014,
and the entire period of CCD observations of 3C~390.3 at the CrAO (1992--2014).
In \tabref{var}, the column ``Bin No.'' is the bin number, the $F_{var}$ parameter
is the rms fractional variability and the $R_{max}$ parameter is simply the
max-to-min ratio of fluxes. Both the parameters are corrected for the
observational uncertainties.

\begin{figure}
  \includegraphics[width=65mm]{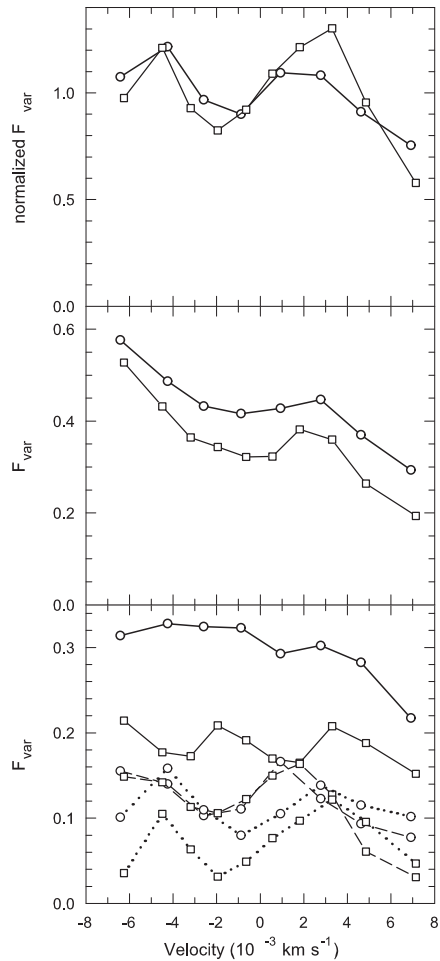}
\caption{Fractional variability amplitudes for the selected bins of the broad line profiles.
  Bottom panel shows amplitudes for the \hb\ (open circles) and \ha\ (open squares)
  lines for 1993--1997 (solid lines), 2005--2007 (dashed lines),
  and 2008--2014 (dotted lines). Middle panel shows amplitudes
  for the entire period of observations, i.e. 1992--2014. Top panel
  shows the mean normalized amplitudes for the three above periods.
  The \hb\ and \ha\ light curves
  are identically sampled in order to correct comparison of amplitudes
  between both lines (see also \tabref{var-sampl}).}
  \label{varfig}
\end{figure}

The considered variability characteristics for any light curve
can only be compared to another light curve when both of them
are sampled identically.
However, there is a notable difference in the sampling of our
\ha-region and \hb-region light curves.
To avoid effect of sampling I have selected quasi-simultaneous data
points from both regions to construct identically sampled light curves.
Their variability characteristics are given in
\tabref{var-sampl} and the $F_{var}$ parameter is shown in \figref{varfig}.

\begin{table*}
  \caption{Variability characteristics of the broad \ha, \hb, and \heii\,$\lambda4686$\,\AA\ emission-lines.}
  \label{var}
  \begin{tabular}{@{}lccccccccccccc@{}}
  \hline
Bin  & & \multicolumn{3}{c}{1993--1997} & \multicolumn{3}{c}{2005--2007} & \multicolumn{3}{c}{2008--2014} &\multicolumn{3}{c}{1992--2014} \\
 No. & Line & Mean & $F_{var}$ & $R_{max}$   & Mean & $F_{var}$ & $R_{max}$ & Mean & $F_{var}$ & $R_{max}$ &  Mean & $F_{var}$ & $R_{max}$ \\
  \hline
1     & \hb & 1.29  & 0.315  & 2.72 & 4.51 &  0.142 & 1.62 & 4.83  &  0.169 & 1.95   & 3.71  & 0.443 & 8.30 \\
2     & \hb & 2.00  & 0.316  & 2.90 & 5.34 &  0.125 & 1.55 & 4.70  &  0.247 & 2.49   & 4.06  & 0.398 & 6.80 \\
3     & \hb & 2.20  & 0.312  & 2.74 & 5.13 &  0.094 & 1.36 & 4.54  &  0.194 & 1.92   & 3.98  & 0.350 & 4.87 \\
4     & \hb & 2.39  & 0.316  & 2.44 & 4.55 &  0.092 & 1.36 & 4.89  &  0.137 & 1.75   & 4.01  & 0.324 & 4.67 \\
5     & \hb & 2.36  & 0.292  & 3.06 & 4.76 &  0.146 & 1.58 & 4.91  &  0.181 & 1.86   & 4.06  & 0.349 & 5.37 \\
6     & \hb & 1.99  & 0.306  & 2.41 & 4.77 &  0.116 & 1.41 & 4.16  &  0.218 & 2.15   & 3.64  & 0.371 & 5.02 \\
7     & \hb & 1.99  & 0.297  & 2.66 & 4.15 &  0.101 & 1.43 & 3.60  &  0.178 & 1.92   & 3.24  & 0.319 & 4.30 \\
8     & \hb & 2.12  & 0.203  & 2.06 & 3.31 &  0.072 & 1.27 & 3.78  &  0.158 & 1.64   & 3.15  & 0.269 & 3.12 \\
1--8  & \hb & 12.7  & 0.308  & 2.56 & 29.6 &  0.112 & 1.43 & 28.0  &  0.188 & 2.00   & 23.5  & 0.366 & 5.18 \\
1     & \ha & 5.46  & 0.199  & 2.14 & 16.5 &  0.159 & 1.50 & 20.4  &  0.052 & 1.34   & 10.4  & 0.601 & 5.92 \\
2     & \ha & 9.18  & 0.163  & 1.90 & 21.3 &  0.155 & 1.50 & 23.9  &  0.119 & 1.59   & 14.0  & 0.471 & 4.08 \\
3     & \ha & 9.09  & 0.159  & 1.78 & 19.1 &  0.123 & 1.38 & 19.7  &  0.081 & 1.48   & 12.7  & 0.390 & 3.14 \\
4     & \ha & 9.10  & 0.189  & 1.94 & 16.4 &  0.111 & 1.31 & 19.1  &  0.051 & 1.34   & 12.2  & 0.364 & 3.10 \\
5     & \ha & 9.09  & 0.172  & 1.86 & 14.0 &  0.129 & 1.30 & 17.2  &  0.063 & 1.41   & 11.2  & 0.335 & 2.85 \\
6     & \ha & 7.67  & 0.156  & 1.55 & 11.1 &  0.159 & 1.44 & 13.8  &  0.096 & 1.66   &  9.07 & 0.330 & 3.21 \\
7     & \ha & 8.34  & 0.161  & 1.62 & 14.7 &  0.177 & 1.58 & 17.7  &  0.115 & 1.69   & 10.9  & 0.399 & 3.20 \\
8     & \ha & 8.58  & 0.189  & 1.84 & 15.8 &  0.138 & 1.41 & 17.5  &  0.130 & 1.64   & 11.3  & 0.379 & 3.30 \\
9     & \ha & 8.11  & 0.180  & 1.84 & 12.9 &  0.066 & 1.23 & 12.9  &  0.107 & 1.55   &  9.58 & 0.274 & 2.50 \\
10    & \ha & 7.45  & 0.150  & 1.70 & 10.2 &  0.033 & 1.20 & 10.9  &  0.054 & 1.26   &  8.63 & 0.207 & 2.07 \\
1--10 & \ha &  83.8 & 0.155  & 1.76 & 155.9&  0.131 & 1.35 & 179.8 &  0.077 & 1.47   & 112.7 & 0.377 & 3.20 \\
$4820$--$4920$\AA & \heii &
              0.858 & 0.373  & 4.48 & 1.68 &  0.253 & 3.05 & 1.19  &  0.539 &$>10$   & 1.24  & 0.470 & $>10$\\

  \hline
\end{tabular}

\smallskip
{\footnotesize
Units for the `Mean' (mean flux) columns are $10^{-14}\,\ergs$.
}
\end{table*}

\begin{table*}
  \caption{Variability characteristics of the identically sampled broad \ha, \hb, and \heii\,$\lambda4686$\,\AA\ emission-lines}
  \label{var-sampl}
  \begin{tabular}{@{}lccccccccccccc@{}}
  \hline
Bin  & & \multicolumn{3}{c}{1993--1997} & \multicolumn{3}{c}{2005--2007} & \multicolumn{3}{c}{2008--2014} &\multicolumn{3}{c}{1992--2014} \\
 No. & Line & Mean & $F_{var}$ & $R_{max}$   & Mean & $F_{var}$ & $R_{max}$ & Mean & $F_{var}$ & $R_{max}$ &  Mean & $F_{var}$ & $R_{max}$ \\
  \hline
1     & \hb & 1.24 & 0.314  & 2.80 & 4.39 & 0.155  & 1.53 & 5.31 &  0.101 & 1.61 & 3.27 & 0.576 & 7.84 \\
2     & \hb & 1.90 & 0.328  & 2.75 & 5.33 & 0.140  & 1.41 & 5.43 &  0.159 & 1.99 & 3.75 & 0.487 & 6.35 \\
3     & \hb & 2.04 & 0.325  & 2.60 & 5.13 & 0.103  & 1.36 & 5.01 &  0.110 & 1.64 & 3.62 & 0.432 & 4.78 \\
4     & \hb & 2.23 & 0.323  & 2.35 & 4.59 & 0.111  & 1.36 & 5.28 &  0.080 & 1.43 & 3.67 & 0.416 & 4.07 \\
5     & \hb & 2.27 & 0.293  & 3.07 & 4.67 & 0.166  & 1.54 & 5.42 &  0.105 & 1.41 & 3.74 & 0.428 & 5.02 \\
6     & \hb & 1.90 & 0.302  & 2.47 & 4.70 & 0.123  & 1.45 & 4.78 &  0.139 & 1.53 & 3.39 & 0.447 & 4.38 \\
7     & \hb & 1.95 & 0.283  & 2.79 & 4.07 & 0.093  & 1.29 & 3.98 &  0.115 & 1.50 & 3.02 & 0.370 & 3.92 \\
8     & \hb & 2.05 & 0.217  & 1.89 & 3.30 & 0.077  & 1.26 & 3.74 &  0.102 & 1.47 & 2.87 & 0.294 & 3.10 \\
1--8  & \hb & 12.1 & 0.309  & 2.54 & 29.4 &  0.123 & 1.36 & 31.2 &  0.112 & 1.59 & 21.4 & 0.453 & 4.86 \\
1     & \ha & 5.57 & 0.214  & 1.86 & 16.7 & 0.149  & 1.53 & 20.5 &  0.036 & 1.21 & 12.9 & 0.528 & 5.13 \\
2     & \ha & 9.33 & 0.177  & 1.71 & 21.5 & 0.142  & 1.46 & 24.2 &  0.105 & 1.38 & 16.7 & 0.432 & 3.70 \\
3     & \ha & 9.16 & 0.173  & 1.66 & 19.3 & 0.113  & 1.36 & 20.0 &  0.063 & 1.24 & 14.7 & 0.364 & 2.86 \\
4     & \ha & 9.19 & 0.209  & 1.75 & 16.5 & 0.106  & 1.32 & 19.2 &  0.031 & 1.17 & 13.9 & 0.344 & 2.74 \\
5     & \ha & 9.21 & 0.191  & 1.69 & 14.1 & 0.122  & 1.32 & 17.3 &  0.049 & 1.37 & 12.6 & 0.322 & 2.57 \\
6     & \ha & 7.68 & 0.170  & 1.63 & 11.2 & 0.150  & 1.48 & 13.9 &  0.077 & 1.34 & 10.2 & 0.323 & 2.85 \\
7     & \ha & 8.30 & 0.165  & 1.60 & 14.8 & 0.164  & 1.59 & 17.9 &  0.097 & 1.39 & 12.6 & 0.382 & 3.21 \\
8     & \ha & 8.66 & 0.208  & 1.78 & 15.9 & 0.128  & 1.44 & 17.7 &  0.122 & 1.65 & 12.9 & 0.359 & 3.21 \\
9     & \ha & 8.14 & 0.188  & 1.94 & 13.0 & 0.061  & 1.25 & 13.1 &  0.096 & 1.32 & 10.6 & 0.264 & 2.66 \\
10    & \ha & 7.64 & 0.152  & 1.65 & 10.2 & 0.031  & 1.21 & 10.9 &  0.047 & 1.17 & 9.22 & 0.194 & 2.20 \\
1--10 & \ha & 84.9 & 0.168  & 1.55 & 157.1& 0.121  & 1.35 & 181.7&  0.060 & 1.28 & 129.1& 0.358 & 2.86 \\
$4820$--$4920$\AA & \heii &
             0.819 & 0.316  & 4.58 & 1.50 & 0.262  & 2.92 & 1.65 &  0.380 & 14.5 & 1.24 & 0.440 & 9.59 \\
  \hline
\end{tabular}

\smallskip
{\footnotesize
Units for the `Mean' (mean flux) columns are $10^{-14}\,\ergs$.
}
\end{table*}

As can be seen from \tabref{var} and \figref{varfig}, the variability
amplitude for 1992--2014 (both $F_{var}$ and $R_{max}$) is the largest
in the far blue wing (bin No.~1) of the \hb\ and \ha\ lines and
it is the smallest in the far red wing of both lines
(bins No.~8 and 10, respectively).
Except for these bins, the variability amplitude is slightly greater
in the blue wing as compared to the red wing.

\subsection{Velocity-resolved lag measurements}
\label{sect-lag}

\begin{figure*}
  \includegraphics[width=140mm]{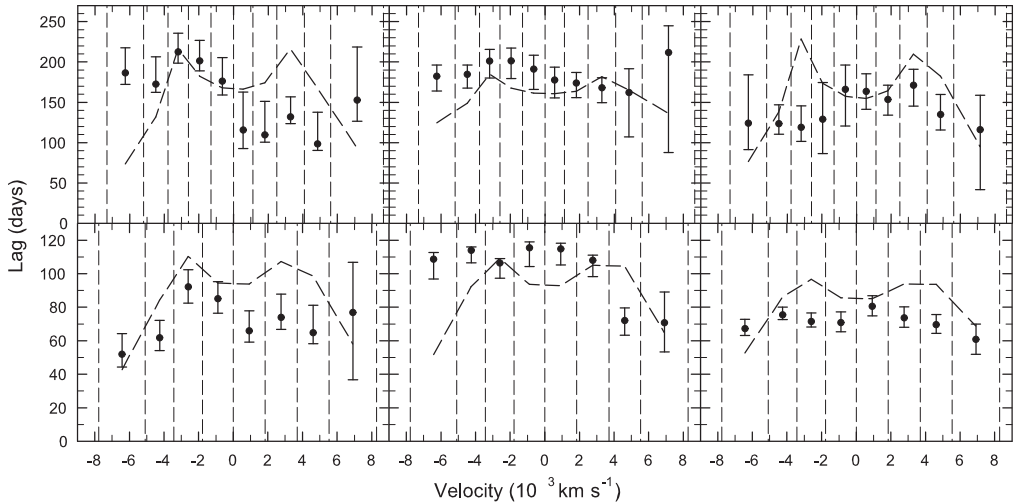}
\caption{Velocity-resolved lag measurements for the \ha\ line (top set of panels)
and for the \hb\ line (bottom set of panels) for the three observational
periods (left to right): 1993--1997, 2005--2007, and 2008--2014. Dashed lines
represent time delays expected from the Keplerian disc model (see text).
Vertical dashed lines are the bin boundaries.}
  \label{lagvel}
\end{figure*}

The time delays between various light curves were determined
as in \citet{Ser02,Ser11,Ser17}, i.e. using the interpolation cross-correlation
function \citep[ICCF, e.g.][]{gas86, wh94}.
The results of the cross-correlation analysis are given in \tabref{tabccf}
and shown in \figref{lagvel}.
for the three periods of observations mentioned above: 1993--1997,
2005--2007, and 2008--2014.

\begin{table*}
  \scriptsize
  \caption{Cross-correlation results.}
  \label{tabccf}
  \begin{tabular}{lccccccccccc}
  \hline
Bin  & Line  & \multicolumn{3}{c}{1993--1997} & \multicolumn{3}{c}{2005--2007} & \multicolumn{3}{c}{2008--2014} & Mean\\
No.  &       & $\tau_{peak}$ & $\tau_{cent}$ & $r_{max}$  & $\tau_{peak}$ & $\tau_{cent}$ & $r_{max}$  &  $\tau_{peak}$ & $\tau_{cent}$ & $r_{max}$  & $\tau_{cent}$\\
\hline
1    & \hb   & $44.9^{+18.3 }_{-3.8 }$  & $51.8^{+12.4}_{-7.6 }$  & 0.932 & $105.2^{+11.7}_{-11.1 }$ & $108.5^{+4.0 }_{-11.7}$ & 0.807 & $57.1^{+11.7}_{-1.4 }$  & $67.1^{+5.6}_{-3.9}$    & 0.911 & $75.8\pm 12.2         $  \\
2    & \hb   & $48.8^{+10.0 }_{-4.9 }$  & $61.6^{+10.5}_{-7.5 }$  & 0.953 & $108.9^{+9.2 }_{-6.2  }$ & $113.8^{+2.0 }_{-7.3 }$ & 0.929 & $57.7^{+10.3}_{-1.0 }$  & $75.3^{+4.7}_{-2.7}$    & 0.966 & $83.6\pm 11.7         $  \\
3    & \hb   & $67.9^{+35.1 }_{-18.1}$  & $92.1^{+10.3}_{-9.8 }$  & 0.926 & $107.3^{+0.8 }_{-13.0 }$ & $106.3^{+2.8 }_{-9.0 }$ & 0.920 & $55.8^{+6.3 }_{-1.6 }$  & $71.3^{+5.1}_{-3.1}$    & 0.948 & $89.9\pm 8.4          $  \\
4    & \hb   & $68.1^{+28.3 }_{-16.7}$  & $84.9^{+10.3}_{-8.5 }$  & 0.936 & $108.3^{+14.8}_{-6.3  }$ & $115.4^{+3.6 }_{-11.1}$ & 0.883 & $55.7^{+23.5}_{-2.4 }$  & $70.7^{+6.4}_{-5.3}$    & 0.921 & $90.3\pm 10.1         $  \\
5    & \hb   & $49.7^{+16.0 }_{-5.0 }$  & $65.8^{+12.0}_{-6.6 }$  & 0.930 & $124.0^{+0.2 }_{-25.0 }$ & $114.7^{+3.5 }_{-9.5 }$ & 0.855 & $54.3^{+66.8}_{-4.1 }$  & $80.4^{+6.4}_{-5.6}$    & 0.910 & $87.0\pm 10.5         $  \\
6    & \hb   & $50.0^{+21.7 }_{-1.8 }$  & $73.8^{+14.0}_{-7.2 }$  & 0.918 & $107.2^{+6.5 }_{-13.0 }$ & $107.9^{+3.2 }_{-9.6 }$ & 0.875 & $54.9^{+30.4}_{-16.0}$  & $73.5^{+6.6}_{-5.5}$    & 0.921 & $85.1\pm 8.7          $  \\
7    & \hb   & $50.3^{+17.6 }_{-1.1 }$  & $64.6^{+16.5}_{-6.4 }$  & 0.918 & $80.1 ^{+5.0 }_{-9.9  }$ & $71.9 ^{+7.5 }_{-8.7 }$ & 0.889 & $54.9^{+3.9 }_{-8.1 }$  & $69.5^{+6.1}_{-5.1}$    & 0.919 & $68.7\pm 5.2          $  \\
8    & \hb   & $40.3^{+106.9}_{-7.6 }$  & $76.8^{+31.5}_{-40.0}$  & 0.825 & $77.2 ^{+14.6}_{-31.1 }$ & $70.6 ^{+18.5}_{-17.3}$ & 0.706 & $54.8^{+14.1}_{-4.9 }$  & $60.7^{+9.3}_{-8.8}$    & 0.508 & $69.4^{+10.9}_{-16.2} $  \\
1    & \ha   & $210.2^{+20.0}_{-14.9 }$ & $186.2^{+31.2}_{-14.0}$ & 0.854 & $182.8^{+35.9}_{-33.1}$  & $182.2^{+13.9}_{-18.1}$ & 0.937 & $113.8^{+71.4}_{-16.5}$ & $124.0^{+59.9}_{-32.6}$ & 0.709 & $164.1\pm 18.2        $  \\
2    & \ha   & $196.2^{+24.7}_{-17.4 }$ & $172.4^{+33.9}_{-10.1}$ & 0.870 & $183.1^{+34.8}_{-12.3}$  & $184.5^{+11.7}_{-16.9}$ & 0.956 & $113.9^{+21.3}_{-6.1 }$ & $123.3^{+23.6}_{-12.8}$ & 0.890 & $160.0\pm 15.0        $  \\
3    & \ha   & $219.9^{+21.2}_{-13.2 }$ & $212.4^{+23.1}_{-13.9}$ & 0.894 & $193.3^{+31.8}_{-9.1 }$  & $200.9^{+14.7}_{-20.9}$ & 0.940 & $116.8^{+19.3}_{-11.8}$ & $118.8^{+26.6}_{-17.5}$ & 0.870 & $177.4\pm 23.2        $  \\
4    & \ha   & $208.2^{+28.5}_{-12.4 }$ & $201.2^{+25.6}_{-12.4}$ & 0.908 & $223.7^{+4.1 }_{-41.4}$  & $201.1^{+16.0}_{-21.8}$ & 0.940 & $116.9^{+64.4}_{-41.7}$ & $128.9^{+45.6}_{-42.5}$ & 0.655 & $177.1\pm 20.2        $  \\
5    & \ha   & $196.0^{+28.1}_{-48.0 }$ & $176.0^{+29.3}_{-17.1}$ & 0.868 & $189.8^{+33.1}_{-29.1}$  & $190.9^{+17.5}_{-24.9}$ & 0.939 & $180.8^{+23.5}_{-67.0}$ & $165.8^{+30.2}_{-45.3}$ & 0.643 & $177.6\pm 16.3        $  \\
6    & \ha   & $91.7 ^{+81.0}_{-11.4 }$ & $115.4^{+47.2}_{-22.9}$ & 0.721 & $182.8^{+24.1}_{-45.0}$  & $177.4^{+16.1}_{-22.1}$ & 0.942 & $181.0^{+6.1 }_{-60.2}$ & $163.2^{+22.3}_{-21.9}$ & 0.816 & $152.0\pm 15.6        $  \\
7    & \ha   & $93.8 ^{+59.0}_{-12.8 }$ & $109.4^{+41.9}_{-8.8 }$ & 0.689 & $182.8^{+17.0}_{-45.7}$  & $173.7^{+13.0}_{-18.0}$ & 0.929 & $135.2^{+45.7}_{-18.1}$ & $153.2^{+17.9}_{-19.1}$ & 0.868 & $145.5\pm 11.8        $  \\
8    & \ha   & $91.7 ^{+83.4}_{-4.4  }$ & $131.8^{+24.9}_{-8.2 }$ & 0.841 & $179.2^{+10.1}_{-45.5}$  & $167.9^{+12.5}_{-18.6}$ & 0.942 & $171.8^{+45.9}_{-52.6}$ & $171.0^{+19.7}_{-25.7}$ & 0.762 & $156.9\pm 11.5        $  \\
9    & \ha   & $76.9 ^{+58.9}_{-8.1  }$ & $98.3 ^{+39.3}_{-7.8 }$ & 0.789 & $188.1^{+8.9 }_{-87.4}$  & $161.9^{+29.7}_{-54.7}$ & 0.851 & $116.9^{+35.0}_{-4.6 }$ & $134.6^{+24.9}_{-19.0}$ & 0.852 & $131.6\pm 17.2        $  \\
10   & \ha   & $195.9^{+40.9}_{-111.7}$ & $152.6^{+65.7}_{-25.9}$ & 0.740 & $197.0^{+52.2}_{-110 }$  & $211.5^{+33.3}_{-124 }$ & 0.698 & $116.8^{+35.5}_{-118 }$ & $115.9^{+42.7}_{-74.3}$ & 0.233 & $160.0\pm 34.2        $  \\
---  & \heii\,$\lambda$4686\,\AA &
               $23.1^{+24.2}_{-13.1}$   & $32.6^{+18.9}_{-13.9}$  & 0.812 & $39.3^{+9.9}_{-7.0}$     & $37.2^{+6.9}_{-7.6}$    & 0.836 & $22.9^{+1.0}_{-14.8}$   & $13.0^{+7.7}_{-8.3}$    & 0.908 & $25.9\pm 8.2          $  \\
\hline
\end{tabular}
\end{table*}

\begin{figure}
  \includegraphics[width=60mm]{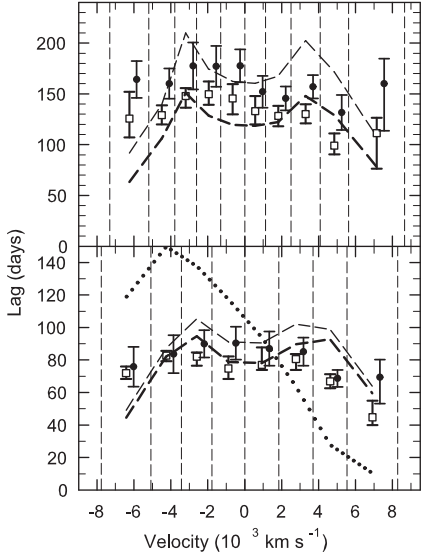}
\caption{Velocity-resolved lag measurements for the \ha\ line (top panel)
and for the \hb\ line (bottom panel). Filled circles are unweighted mean
lag over three periods of observations (lag measurements for each period
are shown in \figref{lagvel}), while
open squares are time delays obtained from the entire observational period
but with removing a long-term trend (see text).
Dashed lines and dotted line
represent time delays as expected from the Keplerian disc model
and from the spherically symmetrical shell with radial infall motion.
The filled circles are offset by $+400$\,\kms\ for clarity.
The thick dashed lines represent best fit to the open squares
by varying the black hole mass in terms of the Keplerian disc model.
Vertical dashed lines are the bin boundaries.}
  \label{lagmean}
\end{figure}

\begin{figure}  \includegraphics[width=60mm]{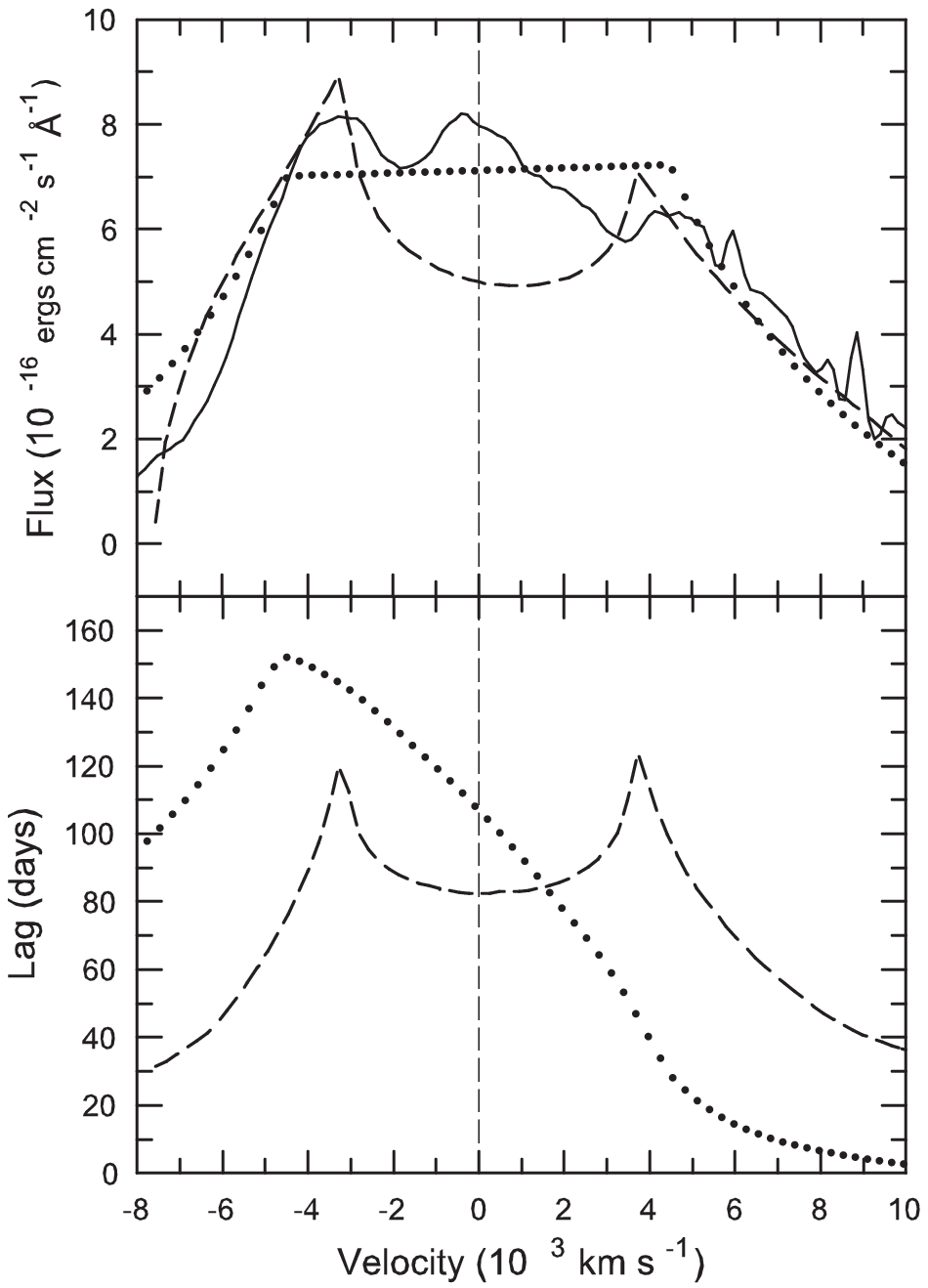}
\caption{Solid line in the top panel is
the observed broad profile of the  \hb\ line. Dashed and dotted lines in the same
panel are fit to the observed profile in terms of the Keplerian disk model
and the spherically symmetric shell with the radial infall motion, respectively.
Bottom panel shows a velocity-resolved lag as expected from both models.}
  \label{lag-mod}
\end{figure}

The lags were measured
from both the location of the maximum value of the ICCF correlation coefficient
($r_{max}$) and from the ICCF centroid based on all points with $r \geq 0.8r_{max}$
(designated as $\tau_{peak}$ and $\tau_{cent}$, respectively).
The lag uncertainties were computed using the model-independent Monte Carlo flux
randomization/random subset selection (FR/RSS) technique described by
\citet{Pet98}. Also given in the last column of Table~\ref{tabccf}
and shown in \figref{lagmean}
is unweighted mean lag for the three periods
of observations. The uncertainties in the mean lag were computed
as described in \citet{Ser17}. Except the bin~\#8 of the \hb\ line,
the derived probability
distributions of the mean lag were found to be close to the normal
distribution, so only $\pm$ uncertainties are given
(no $+$ and $-$ separately).

As was shown in \citet{Ser17} the reverberation mapping can only be applied to the entire
period of observations of the 3C~390.3 nucleus after
removing a long-term trend. This trend has been expressed by a slowly varying
scalefactor $c(t)$ in the power-law relationship
between the line and continuum fluxes:  $F_{line}\propto c(t)\,F_{cont}^a$.
In \citet{Ser17}, a power-law index $a$ was found to be $0.77$ and $0.54$
for the \hb\ and \ha\ total fluxes, respectively. This lag has been computed
for the considered bins and it is shown in \figref{lagmean}.
As can be seen from \figref{lagmean}, the lag for the entire period
is slightly less than the lag averaged over the individual periods
(the lag from \tabref{tabccf}, last column).

Finally, it was checked what the velocity-dependent lag is expected
from the simplest kinematic models of the BLR.
In \citet{Ser02} a model profile produced by the relativistic Keplerian disc
configuration \citep[e.g.][]{ch89b, er94} has been fitted to the mean observed
profiles of the \hb\ and \ha\ lines. For the \hb\ line, the following disc parameters
have been obtained: $r_{in} = 104\,r_g$,
$r_{out} = 840\,r_g$, $i=27\degr.6$, and $\alpha = 2.48$, where $r_g = 2\,G\,M /
c^2$ is the gravitational radius and the disc parameters are the inner/outer
radii, inclination angle ($i=0$ for pole-on configuration), and emissivity
per unit surface: $\varepsilon(r) \propto r^{-\alpha}$. \figref{lag-mod} shows
the observed and fitted profiles for this configuration as well as
the lag--velocity dependence expected from.
To simulate observed \hb\ and \ha\ light curves, the model-dependent
mass of the central
black hole has been set to $2.5\times 10^9\,M_\odot$ and $4.1\times 10^9\,M_\odot$
for the \hb\ and \ha\ lines, respectively
\citep[see Sect.~3.5, Table~7 from][]{Ser17}. The mass difference
probably reflects a difference between kinematics of the emission regions of both lines.
The expected lag--velocity dependence for the Keplerian disc model is shown
in \figref{lagvel} and \ref{lagmean} by dashed lines. The thick dashed line
in \figref{lagmean}
represents the best fit to the observed lag--velocity dependence
by varying the black hole mass in terms of the Keplerian disc model.
The best fit gives the black hole mass of
$2.4\times 10^9\,M_\odot$ and $3.1\times 10^9\,M_\odot$
for the \hb\ and \ha\ lines, respectively.

Similarly, I have considered
a spherically symmetric shell with radial free fall motion. The fitting results
for the \hb\ line
and the expected lag--velocity dependence for this configuration
are shown in the \figref{lagmean}.
For the \hb\ line the following shell parameters
have been obtained: $r_{in} = 510\,r_g$,
$r_{out} = 4450\,r_g$, and $\alpha = 3.22$, where emissivity
per unit volume is $\varepsilon(r) \propto r^{-\alpha}$.
The model-dependent black hole mass has been selected to match
the observed lag for the
total \hb\ fluxes and it was found to be $5.0\times 10^8\,M_\odot$.

It is obvious that the observed velocity-dependent lag
strongly contradicts the spherical infall configuration for which
a maximum lag value among the bins in the blue wing should exceed
a minimum lag value among the bins in the red wing by 15 times
(\figref{lagmean}).
Instead, the observed dependence is similar to that expected from the
disc configuration as can be seen from \figref{lagvel} and \ref{lagmean},
although there are some differences. E.g. the observed lag for the bin~\#7
for the \hb\ line is significantly less than the expected from the disk model
and the observed lag--velocity dependence does not look clearly
a double-peaked as expected. Also, it seems that the blue wing lag is
greater than the red wing lag, that is a signature of infall motion.
However, the significance of the lag difference between both wings as compared
to the expected difference in terms of the considered disc model
was found to be $\approx 2.5\sigma$, so the presence of an additional infall motion
is not significant enough.

There are numerous observational evidences for the existence
of the gas infall motion in AGNs, e.g.
reverberation mapping results \citep{Den09,Gri13,Gri17} or redshifted absorption lines
\citep{Shi16,Rub2017,Zhou2019}. It is obvious that there must be mechanisms for
angular momentum loss of the line-emitting gas for such a motion.
These mechanisms (except for galaxy interactions) can, for example, be the
tidally disrupted dusty clumps \citep{Wang2017}
or magneto-rotational instability \citep{Gas13}.

\subsection{\heii\,$\lambda4686$\,\AA}
\label{sect-heii}

Since the broad \heii\,$\lambda4686$\,\AA\ line is blended with the
broad \hb\ line, it is impossible to measure its total flux. So, its
flux has been measured in the blue wing at the wavelength range
of 4820--4920\,\AA\ (heliocentric reference frame), that is at the
line-of-sight velocity range from $-7730$ to $-1570$\,\kms.
It seems that the broad \heii\
line is free from any other broad-line features at above range.
Several weak narrow-line features were removed from the
broad \heii\ profile using the \oiii\,$\lambda$5007\,\AA\ narrow-line profile
as a template in order to obtain a pure \heii\ broad-line fluxes.
The continuum zones for the \heii\ line were chosen to be
4428--4448\AA\ and 4735--4775\AA. The narrow \heii\
line fluxes have been measured as well using the \oiii\ line profile
as a template to separate \heii\ narrow line from the \heii$+$\hb\ broad
line profiles.

\begin{figure}
  \includegraphics[width=\columnwidth]{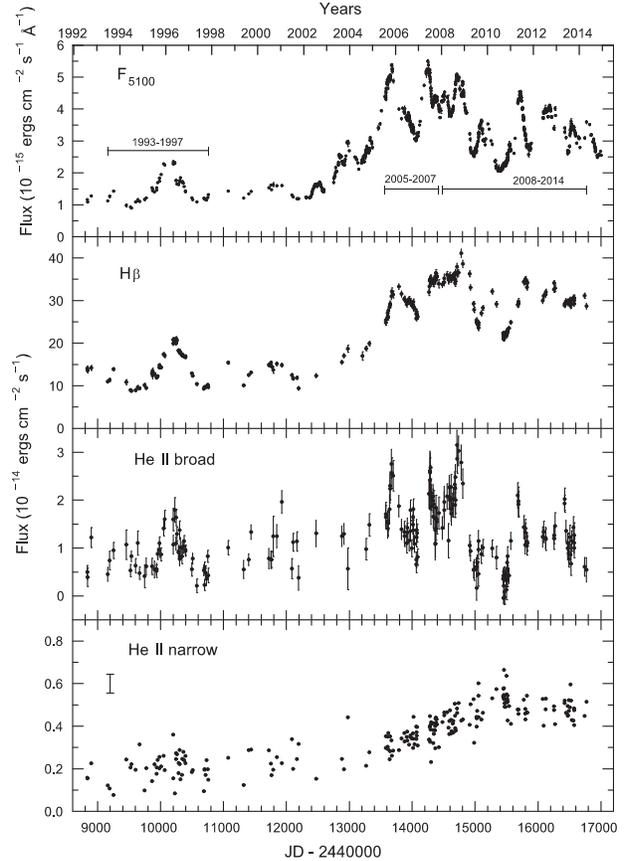}
\caption{Light curves (top to bottom) for the continuum at
$\lambda$5100\AA\ and for the broad \hb\ and broad
and narrow \heii\,$\lambda4686$\,\AA\ emission lines. The vertical
error bar in the bottom panel represents uncertainties derived from
the point-to-point scattering.}
  \label{lc-heii}
\end{figure}

Both the broad and narrow \heii\ line light curves are shown
in \figref{lc-heii}. The optical continuum and the \hb\ light curves
from \citet{Ser17}
are shown in the same figure for comparison.
The variability characteristics of the broad \heii\ line are given in \tabref{var}
and \ref{var-sampl}, last rows.
Both the $F_{var}$ and $R_{max}$ values are corrected
for the observational uncertainties.
As can be seen from \tabref{var} and \ref{var-sampl} the variability
amplitude of the broad \heii\ line is greater than that of the Balmer lines
for all observational periods. However,
the \heii\ fractional variability amplitude ($F_{var}$)
for the {\em entire} observational period is comparable to that of \hb\
(\tabref{var-sampl}), while the
max-to-min ratio of fluxes ($R_{max}$) is much greater for the \heii\ line.
This is because the Balmer lines and the optical continuum contain
a long-term trend that is not seen in the helium line
(\figref{lc-heii}).
This trend provides
an additional increase in the variability amplitude of the Balmer lines
and the optical continuum as compared to the \heii\ line
when the entire observational period is considered.

\begin{figure}
  \includegraphics[width=0.78\columnwidth]{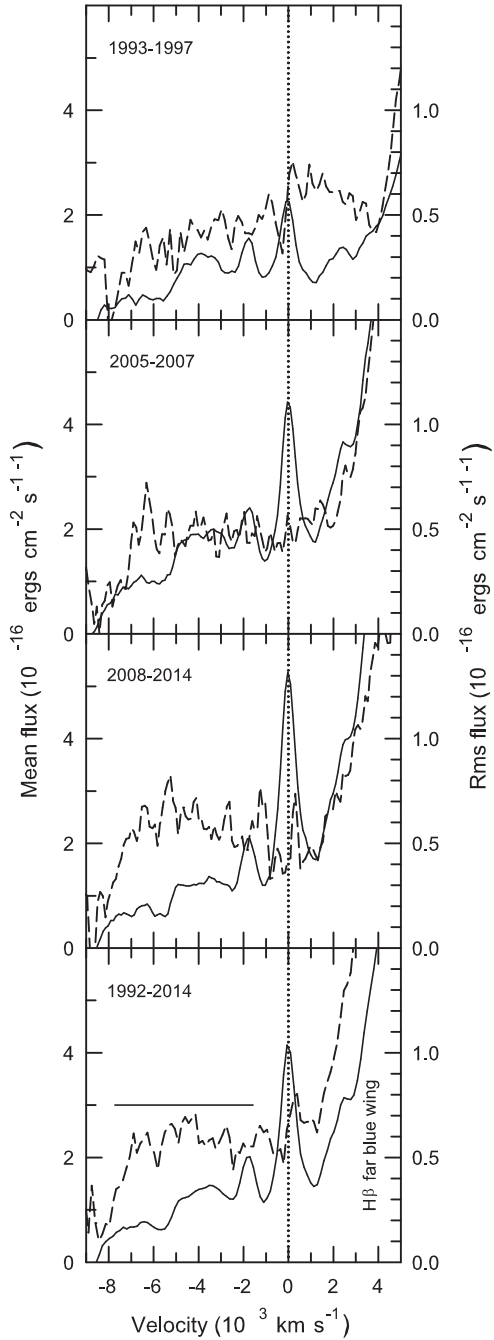}
\caption{
Mean (solid lines) and rms (dashed lines) profiles of the
\heii\,$\lambda4686$\,\AA\ line superposed on the far blue wing of the broad \hb\ line
for the following periods of observations (top-to-bottom): 1993--1997, 2005--2007, 2008--2014,
and 1992--2014. The left-hand axes are the mean fluxes, while the right-hand
axes are the rms fluxes. The rms pfofiles consist the broad-line components
only (narrow-line components are previously subtracted), while the mean profiles
consist both the broad and narrow components. Vertical dotted lines indicate
zero velocity, while the solid horizontal line at the bottom panel indicates
the intergration zone to obtain the light curve of the broad \heii\ line.
The variability of the narrow \heii\ line is clearly seen.}
  \label{rms-heii}
\end{figure}

The narrow \heii\ line is variable as well with
the variability amplitude $R_{max}\approx 3$. It is greater than the variability
amplitudes of both the narrow Balmer lines ($R_{max} = 1.6$) and
narrow \oiii\ line ($R_{max} = 1.3$), see \citet{Ser17}.
The cross-correlation function for the narrow \heii\ light curve
gives a response to the continuum variations with a lag
of $\tau_{cent} = 1298^{+90}_{-78}$\,d. and $r_{max} = 0.825$.
However, cross-correlation analysis can not be applicable in this case because
the narrow \heii\ light curve is not only time-shifted, but too smooth.
It does not trace continuum events except for the long-term trend.

\begin{figure}
  \includegraphics[width=60mm]{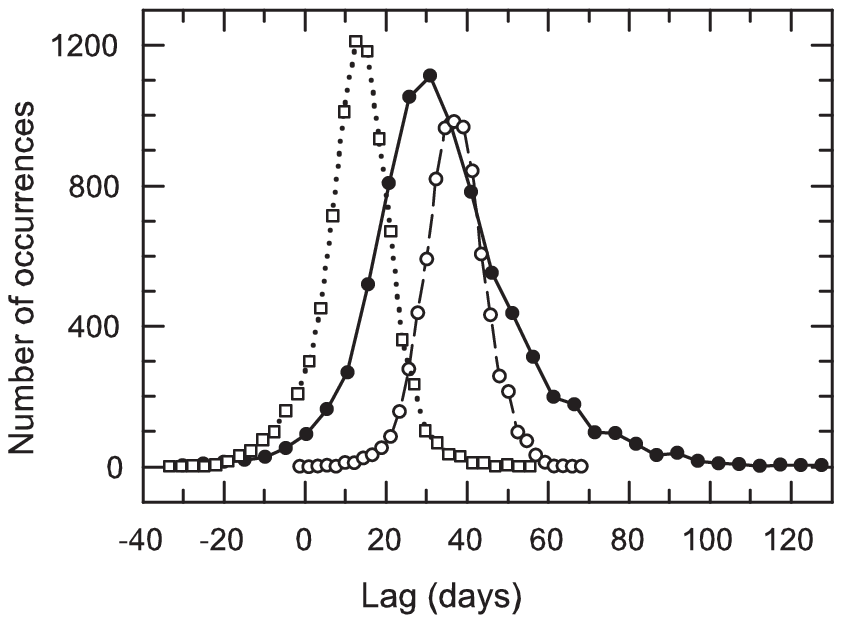}
\caption{Probability distributions for a lag ($\tau_{cent}$) between
  optical continuum  and \heii\,$\lambda 4686$\,\AA\ broad emission line
  for 1993--1997 (filled circles connected by a solid line), 2005--2007
 (open circles connected by a dashed line), and 2008--2014
 (open squares connected by a dotted line). The distributions are derived
  from the FR/RSS technique.}
  \label{hst-heii}
\end{figure}

The cross-correlation results for the broad \heii\ line are given in \tabref{tabccf},
last row.
As can be seen from \tabref{tabccf}, the lag uncertainty for the \heii\ line
for 1993--1997 is more than twice greater than the uncertainties for
2005--2007 and 2008--2014, while the uncertainties for the last two periods
are approximately the same.  Therefore the mean lag value in the last colum
and last row of \tabref{tabccf} is a weighted mean.
The statistical weights were chosen to be 1/4, 1, and 1 for above three periods.
The uncertainty in the mean lag value was computed as described in \citet{Ser17}
using probability distributions for $\tau_{cent}$ for each period.
These distributions were obtained using the model-independent Monte Carlo flux
randomization/random subset selection (FR/RSS) technique described by
\citet{Pet98} and they are shown in \figref{hst-heii}.
The mean lag value for the \heii\ line is $26\pm 8$\,d, significantly less
than that of the Balmer lines \citep[cf.][]{Ser17},
so the \heii\ emission region is much smaller in size.

Finally, I have applied the cross-correlation analysis for the \heii\ line
to the entire observational period, but with
removing a long-term trend. As for the Balmer lines
in \citet{Ser17}, this trend has been expressed by a slowly varying
scalefactor $c(t)$ in the power-law relationship
between the line and continuum fluxes:  $F_{line}\propto c(t)\,F_{cont}^a$.
The lag was found to be $25.9^{+7.2}_{-9.0}$\,d, while
a power-law index $a$ equals to $1.03$. In \citet{Ser17}
this index was found to be $0.77$ and $0.54$
for the \hb\ and \ha\ total fluxes, respectively.
So, the variability amplitude must be largest in the \heii\ line,
less in \hb, and more less in \ha, just as observed.

\section[]{Summary}
\label{sect-sum}

\begin{enumerate}


\item
The observed velocity-dependent lag for the Balmer lines
is similar to that expected from the Keplerian disc configuration,
although there are some differences.
Probably, a radial infall motion can be present in the BLR of 3C~390.3
in addition to the Keplerian rotation.

\item
The mean weighted lag of the broad \heii\,$\lambda$4686\AA\ emission line
is $26\pm 8$\,d,
significantly less than that of the Balmer lines \citep[cf.][]{Ser17},
so the \heii\ emission
region is much smaller in size. This value need to be corrected
for time dilation by dividing by $1+z$ to put it into the rest frame.

\item
In terms of the power-law relationship between line and continuum fluxes
in 3C~390.3:
$F_{line}\propto c(t)\,F_{cont}^a$, where $c(t)$ is a
slowly varying scalefactor, a power-law index $a$
for the broad \heii\,$\lambda$4686\AA\ line
is equal to $1.03$, while according to \cite{Ser17}
it is equal to $0.77$ and $0.54$ for the \hb\ and \ha\ lines, respectively.
It means that the variability amplitude is
the largest in the \heii, less in \hb, and more less in \ha.
However, the \heii\ variability amplitude ($F_{var}$)
for the {\em entire} observational period is comparable to that of \hb\
because the Balmer lines and optical continuum light curves
both contain a long-term trend that is not seen in the helium line.

\item
The narrow \heii\,$\lambda$4686\AA\ line is variable with
the variability amplitude $R_{max}\approx 3$. It is greater than the variability
amplitudes of both the narrow Balmer lines ($R_{max} = 1.6$) and
the narrow \oiii\,$\lambda$5007\,\AA\ line ($R_{max} = 1.3$), see \citet{Ser17}. So, the narrow
\heii\ line emission region is more compact than that of the Balmer lines and \oiii.

\item
The variability amplitude is found to be
the largest in the far blue wing of the broad \hb\ and \ha\ lines and
it is the smallest in the far red wing of both lines.

\end{enumerate}

\section*{Acknowledgments}

The CrAO CCD cameras have been purchased through the US Civilian Research
and Development Foundation for the Independent States of the Former Soviet
Union (CRDF) awards UP1-2116 and UP1-2549-CR-03.

\label{lastpage}

\end{document}